 \documentclass[twocolumn, a4paper, 10pt, showpacs, prb, reprint, aps,superscriptaddress, amsmath, amssymb, amsfonts, floatfix]{revtex4-1}

\usepackage{amsmath, amssymb, latexsym}
\usepackage{amsfonts}
\usepackage{grffile}
\usepackage{graphicx}
\usepackage{dcolumn}
\usepackage{xcolor}
\DeclareMathOperator{\Tr}{Tr}

\begin{document}
	
	\title{Power-law energy level-spacing distributions in fractals}
	
	\author{Askar A. Iliasov}
	\email{A.Iliasov@science.ru.nl}
	\affiliation{Institute for Molecules and Materials, Radboud University, Heyendaalseweg 135, 6525AJ Nijmegen, The Netherlands}
	
	\author{Mikhail I. Katsnelson}
	\affiliation{Institute for Molecules and Materials, Radboud University, Heyendaalseweg 135, 6525AJ Nijmegen, The Netherlands}
	
	\author{Shengjun Yuan}
	\email{s.yuan@whu.edu.cn}
	\affiliation{School of Physics and Technology, Wuhan University, Wuhan 430072, China}
	\affiliation{Institute for Molecules and Materials, Radboud University, Heyendaalseweg 135, 6525AJ Nijmegen, The Netherlands}
	
	\date{\today}
	
	\begin{abstract}
		In this article we investigate the energy spectrum statistics of fractals at the quantum level. We show that the energy-level distribution of a fractal follows a power-law behaviour, if its energy spectrum is a limit set of piece-wise linear functions. We propose that such a behaviour is a general feature of fractals, which can not be described properly by random matrix theory. Several other arguments for the power-law behaviour of the energy level-spacing distributions are proposed.
	\end{abstract}

	
	\maketitle

	\section{Introduction}
	
	The quantum chaos theory successfully describes mesoscopic systems with the help of random matrix theory \cite{Dyson1962,Peerenboom1981,Halperin1986,Stockmann_2000,Bohr_1969,Beenakker1997,Mehta_2004}.
	 Prominent class of such systems are disordered systems. In general, they could have some properties of fractals, such as the fractional dimension and the singular spectrum \cite{HavAbr1987,EversMirlin2008}. Surprisingly, much less is known on energy spectrum of regular fractal structures, probably, due to a lack of motivation: whereas disordered systems are very common in physics, regular fractals, small enough to make quantum effects relevant, were considered as exotic. The situation was changed just recently due to the developing of new nanofabrication methods \cite{Polini_etal2013, Gibertini_etal2009, Shang_etal2015}. As an example, self-similar structures are already applied in the production of antennas and metamaterials \cite{Hohlfeld2011, Huang2010}.
	
Fractals were intensively studied in physical \cite{HavAbr1987,Tosatti_1986,Feder_1988} and mathematical \cite{Strichartz2006} literatures. However, quantum effects are hardly studied; the previous researches did not use the language of quantum chaos or random matrix theory. Recent works along this direction include topological characteristics of fractals \cite{Brzezetal2018}, localization in randomly generated fractals \cite{KosKrzys2017}, and the transport \cite{Veen2016}, optical \cite{Veen2017} and plasmonic \cite{Westerhout2018} properties of  regular fractal structures such as Sierpinski carpet and Sierpinski gasket.

The statistics of spectrum of Sierpinski carpet and random carpet with the same number of holes were studied in Ref. {[}\onlinecite{Hernando2015}]. It was shown that the spectrum statistics of these two types of carpets are drastically different. The random carpet has usual Poisson statistics of the energy spectrum while Sierpinski carpet demonstrates a power-law distribution. The power-law behaviour also was numerically demonstrated for Sierpinski gasket with disorder \cite{KatEvan1996}.
	
For fractals with finite ramification number, the spectrum of such system is a limit set of a map, which is inverse to polynomial. This procedure is called spectral decimation \cite{Teplyaev_2007, Sabot_2003} and was first applied in the case of Sierpiski gasket \cite{Domany_1983}. The ramification number of a fractal is given by the number of bonds that need to be cut to separate  two iterations from the another. The spectral decimation is the procedure which relates iterations of the fractal with iterations of some functions.
To our knowledge, the level-spacing distribution was not analytically studied even for Sierpiski gasket or other fractals which admit spectral decimation.
	
 Our paper is devoted to the analysis of quantum energy spectrum of some fractal structures. Section II describes the spectrum statistics of Sierpinski gasket and argues that the power-law distribution could be the feature of other fractals admitting spectral decimation. Section III demonstrates the spectrum statistics for modified Serpienski gasket, and Section IV provides a detailed discussion of the results.
	
	\section{Sierpinski gasket}	
	
	\subsection{Symmetries of spectrum}
	
	We use the following simplest, single-orbital tight-binding Hamiltonian to study the energy spectrum of fractal structures:
	
	\begin{equation}
	H= -t\sum_{\langle ij\rangle} c^\dag_i c_j \,,
	\label{Eq:TBmodel}
	\end{equation}
	
where $\langle ij\rangle$ denotes the nearest-neighbor sites belonging to the studied fractal.
	
\begin{figure*}[ht!]
		\mbox{
		\includegraphics[width=0.25\linewidth]{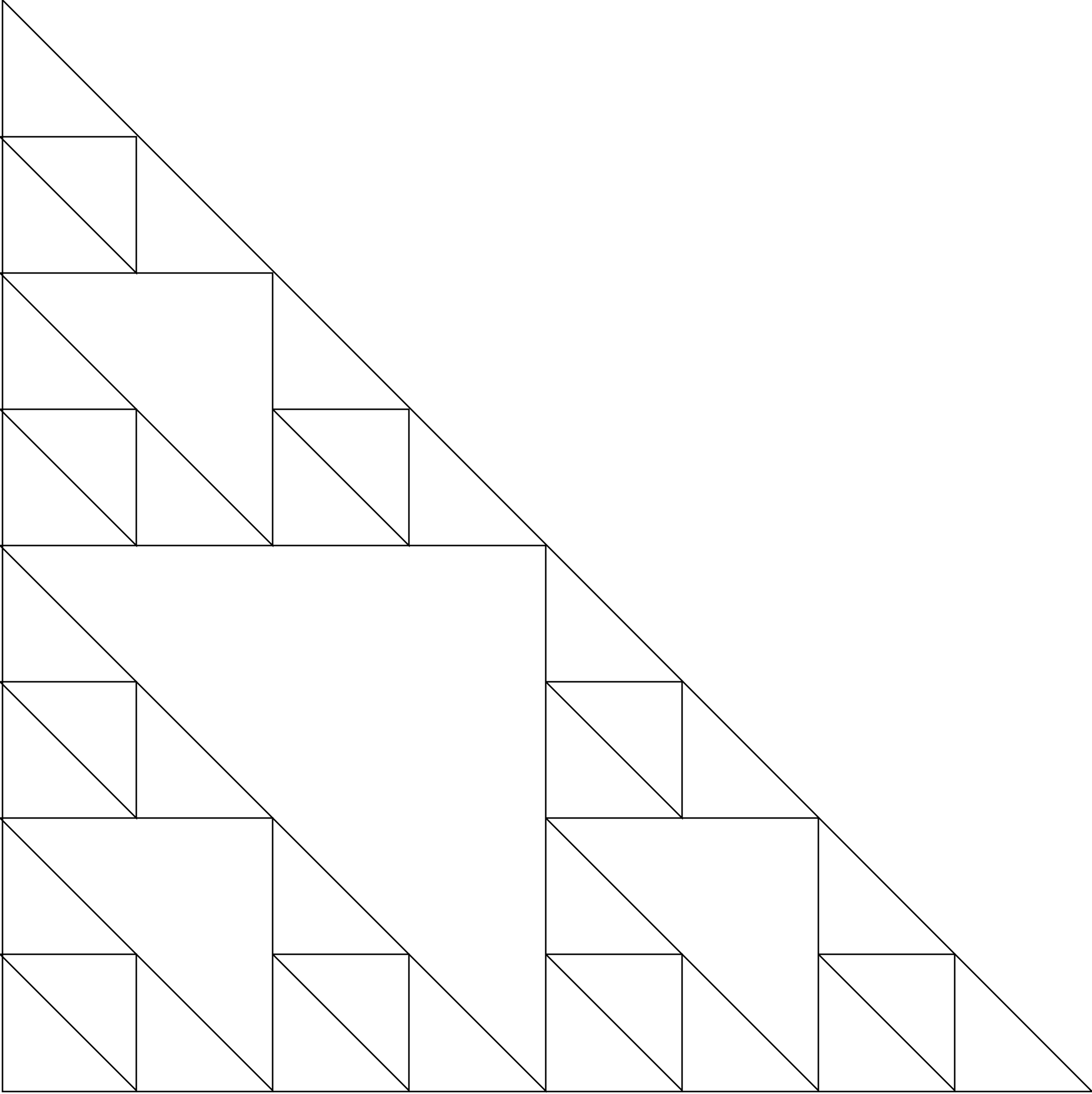}
		\includegraphics[width=0.425\linewidth]{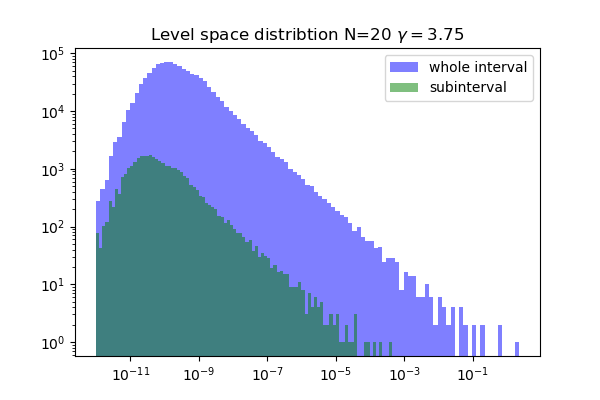}
		}
		\caption{		\label{Fig:Logistic_LSD_twofigures} Left: The Sierpinski gasket with 3 iterations. Right: The level-spacing distribution of Sierpinski gasket for 20 iterations (blue
			color) and left subinterval consisting of 15 iterations (green color).}

	\end{figure*}

In the case of Sierpinski gasket (an example of $n=3$ iterations is shown in the Fig. \ref{Fig:Logistic_LSD_twofigures}), the energy spectrum is generated by the following functions \cite{Domany_1983}:
	
	\begin{equation}
	x_{n+1}=F_{\pm}(x_n)=\pm \sqrt{\gamma-x_n}\,,
	\label{Eq:Dynsys}
	\end{equation}
with $\gamma=15/4$ and the variable ${x_n}$ corresponding to the spectrum of the tight-binding model in Eq. (\ref{Eq:TBmodel}) (in the units of $t$). These functions produce the spectrum of $n+1$-th iterations of Sierpinski gasket from the spectrum of $n$-th iterations. There is a shift in the functions of Eq. (\ref{Eq:Dynsys}) $x \rightarrow x+3/2$ in comparison with Ref. [\onlinecite{Domany_1983}] (see their Eq. (2.16)), for the sake of convenience.
	
The spectrum of Sierpinski gasket consists of a limit set of the dynamical systems of Eq. (\ref{Eq:Dynsys}) and a non-regular part of degenerate eigenvalues. However, the number of degenerate eigenvalues are much smaller for large enough iterations of the fractal, and in the following discussion we do not take them into account.
	
	To simplify the following analysis, let us describe some simple properties of Eq. (\ref{Eq:Dynsys}), following the discussion in Ref. [\onlinecite{Domany_1983}]. One can see, that $F^{-1}_{+}(x)=F^{-1}_{-}(x)=x^2-\gamma$, which is a polynomial map closely connected to logistic map $r x(1-x)$, which is intensively studied in chaotic dynamics \cite{Strogatz_2000}. The limit set $K=\lim_{n\to\infty} \cup F_{\pm}\circ F_{\pm}\circ \ldots \circ F_{\pm}(x_0)$ is Julia set of the real polynomial map, i.e. the real part of Mandelbrot set. $K$ is bounded by the values $x_{\max}$ and $x_{\min}=-x_{\max}$, which are determined by equation $x_{\max}=\sqrt{\gamma+x_{\max}}$. In the case of Sierpinski gasket, $\gamma=15/4$ and $x_{\max}=2.5$.
	
	The set $K$ is invariant under the action of $F_{\pm}$ and $F^{-1}_{\pm}$. Therefore, $K$ is invariant under the action of compositions of $F_{\pm}$, i.e., sequences $F_{\pm}\circ F_{\pm}\circ \ldots \circ F_{\pm}$. Let us denote such a composition $F^{n}_{\alpha}$, where the index $n$ is the number of iterations, and $\alpha$ is a sequence of $"+"$ and $"-"$ for each iteration. Every $F^{n}_{\alpha}$ is a monotonic function of $n$, since its compositions are monotonic. Therefore $F^{n}_{\alpha}$ maps the interval $I_{\max}=[x_{\min},x_{\max}]$ to some interval $I^n_{\alpha} \in I_{\max}$. By the same arguments $I^{n+k}_{\alpha\alpha'} \in I^n_{\alpha}$. Thus one can see the self-similar structure of the spectrum, since every interval $I^n_{\alpha}$ contains the set $K$, deformed by the monotonic function $F^{n}_{\alpha}$.
	
	One can visualize the dynamics of intervals in the following way. After $n$ iterations of $F_{\pm}$, the initial interval is divided into $2^{n}$ disjoint subintervals. The set of the subintervals have hierarchy induced by the relation $I^{n+k}_{\alpha\alpha'} \in I^n_{\alpha}$. The hierarchy allows to introduce a natural order on the subintervals. Let $\alpha$ be some string of $n$ symbols $"+"$ and $"-"$. A string of symbols $+$ and $-$ can be added to another string by concatenation (for example, if $\alpha$ is a string $"+-"$, then $"-+"\alpha="-++-"$, $"-"\alpha="-+-"$ and so on). One can deduce the order by iterations: if $\alpha_1<\alpha_2$, then $"-"\alpha_1<"-"\alpha_2$ and $"+"\alpha_1<"+"\alpha_2$, i.e. $"-"$ sign keeps the order, $"+"$ sign inverts it. The closest strings are different only in one sign and changes of sign with $\alpha$ increasing occurs in the same positions as in usual numbers. For example, if $n=3$, the order is $\{"---", "--+", "-++", "-+-", "++-", "+++", "+-+", "+--"\}$. Thus one can estimate the location of interval $I_{\alpha}$.
	
	The derivative of $F^{n}_{\alpha}$ with respect to $x$ goes to zero with $n$ increasing, therefore if $n$ is large enough, $F^{n}_{\alpha}$ is almost a constant function on $I_{\max}$ and independent of $x_0 \in I_{\max}$.
	The derivative of $F^{n}_{\alpha}$ is obtained by the product of derivatives $F'_{\pm}$ (prime means $d/dx$) calculated in the points of the corresponding sequence ${x_l}$, where $x_l=F^l_{\alpha_l}(x_0)$:
	
	\begin{equation}
	(F^{n}_{\alpha})'(x_0)=\Pi_{l} F'_{\pm}(x_l)
	\end{equation}
	
	If one wants to consider one of the subintervals, one should add finite number of $F_{\pm}$ to all possible $F^{n}_{\alpha}$. This is the same as if one considers only $\alpha$ beginning with a particular string. Since additional $F_{\pm}$ do not effect on the corresponding sequences ${x_l}$, the level-spacing distribution on a subinterval would be just multiplied by a constant (a shift in log-log coordinates) in comparison with the whole interval. This effect is demonstrated in the Fig. \ref{Fig:Logistic_LSD_twofigures}.

	To investigate the structure of the level-spacing distribution, let us first consider a toy model.
	
	\subsection{Toy model}
	
\begin{figure}[ht!]
		\includegraphics[width=1.05\linewidth]{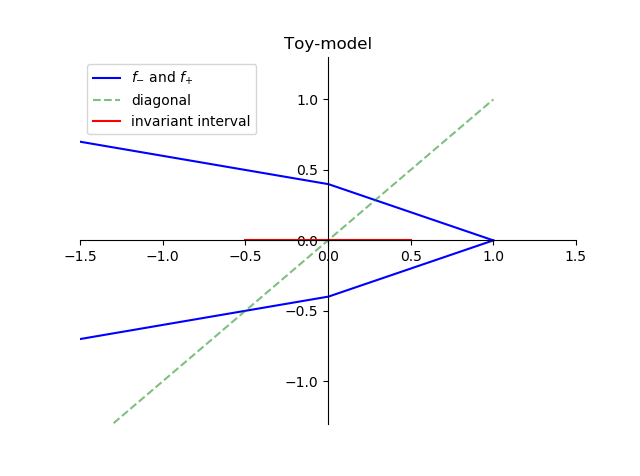}
		\caption{\label{Fig:toymodel} Toy model of dynamical system. The dynamical system is represented by two branches of piece-linear functions. New points are obtained from previous ones by these functions.}		
	\end{figure}

	A toy model can be regularly constructed as a piece-wise linear approximation of the dynamical system described by equation (\ref{Eq:Dynsys}). There are two continuous functions $f_+$ and $f_-$, which are symmetric with respect to zero $f_-(x)=-f_+(x)$. Each of the functions have two segments, when $x<0$ and $x>0$. An example of such dynamical system is shown in Fig. \ref{Fig:toymodel}. One can see that all discussion of previous section is also applied to this model.
	
	The system has $4$ parameters $\alpha$, $\beta$, $\gamma_{\alpha}$, $\gamma_{\beta}$ describing linear functions:
	
	\begin{equation*}\label{Eq:toymodel}
	f_+(x)=\begin{cases}-\beta(x-\gamma_\beta),&\mbox{if } x<0\\ -\alpha(x-\gamma_\alpha),&\mbox{if } x>0 \end{cases}
	\end{equation*}

	There is an obvious relation $\alpha\gamma_\alpha=\beta\gamma_\beta$. The system acts on an invariant interval $I_{\max} =[-x_{\max},x_{\max}]$, where $x_{\max}=\gamma_{\beta}\beta/(1-\beta)$.
	
	After the first iteration of $f_{\pm}$ on invariant interval, the image consists of two disjoint segments $I_+=f_+(I_{\max})$ and $I_-=f_-(I_{\max})$, in other words, a gap $\Delta_0$ occurs in the interval. After the second iteration, $\Delta_0$ is mapped into two symmetric gaps $\Delta_1$ in $I_+$ and $I_-$. Therefore, the following iterations produce new gaps only by linear transformations.
	The gap lengths after the second iterations are $\alpha|\Delta_1|$ and $\beta|\Delta_1|$.
	
	Since the dynamical system is symmetric, each of new gap lengths are obtained by both multiplications $\alpha$ and $\beta$. Thus the new lengths after $n$-th iteration are distributed binomially.
	
	\begin{equation}
	p_n(s)=2\sum^{n-1}_{k=0} {n-1\choose k} \delta(s-\alpha^k\beta^{n-k}|\Delta_1|)
	\end{equation}
	
	The full distribution of all gap lengths after $n$ iterations $P_n (s)$ is just sum of $p_n(s)$.
	
	If $\alpha=\beta$, $p_n(s)$ becomes just a delta function $p_n(s)=2^{n}\delta(s-\beta^{n-1}|\Delta_1|)$ and $P_n (s)$ becomes a power law $P_n (s) \sim s^{\ln2/\ln\beta}$. It should be noted that the estimation is correct only if $\beta<0.5$, otherwise there will be no gaps and the limit set of the dynamical system coincides with the interval.
	
\begin{figure*}[ht!]
		\mbox{
		\includegraphics[width=0.4\linewidth]{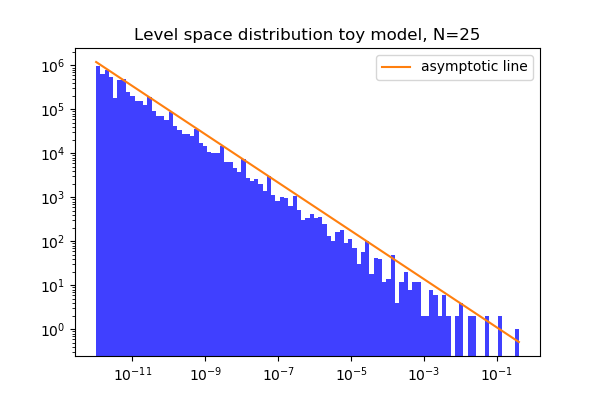}
		\includegraphics[width=0.4\linewidth]{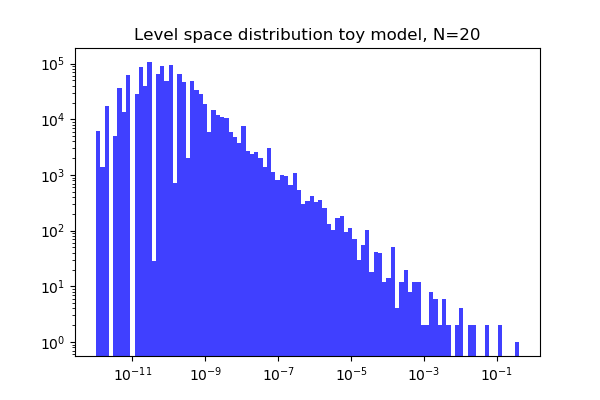}
		}
		\caption{\label{Fig:toymodelLSD_twofigures} Left: The level spacing distribution for the toy model in Fig. \ref{Fig:toymodel}, N=$25$ iterations. Asymptotic line is $P(s)=(\frac{s}{|\Delta_1|})^{\frac{2\ln 2}{\ln\alpha+\ln\beta}}$. Right: The level spacing distribution for the toy model in Fig. \ref{Fig:toymodel}, N=$20$ iterations.}

	\end{figure*}

	Consequently, $p_n(s)$ can be seen as smearing of delta-peak with maxima at $\alpha^{\frac{n}{2}}\beta^{\frac{n}{2}}$, asymptotic value $2^n$ and support $S_n= [|\Delta_1|\beta^n,|\Delta_1|\alpha^n]$. Since $\beta<\alpha<0.5$, these $S_n$ do not intersect with each other starting from some $n_0$. So, the asymptotic  for $P_n(s)$ is the following: $P_n (s) \sim s^{\ln2/(\frac{\ln\beta+\ln\alpha}{2})}$ (Fig. \ref{Fig:toymodelLSD_twofigures}).

	\subsection{Level-spacing distribution}
	
	If we add additional line segments into the described toy model so that there are $k$ slopes $\beta_j$, the model changes slightly. The smearing widths of delta peaks do not change, since they depend only on maximal and minimal slopes. The total number of gaps on the $n$th iteration remains the same $2^n$. We have to change only the exponent in the asymptotic power law, which will be equal to the mean value of the logarithms of slopes, that is, $\frac{1}{k}\sum \ln \beta_j$. The level-spacing distribution for Sierpinski gasket is the limit of such systems, so asymptotic exponent is described by the mean of derivatives of $F_{\pm}$ on the limit set $K$ of the dynamical system.
	
	However, the spectrum of $n$ iterations of a fractal corresponds to $n$ iterations of dynamical system starting from a few points, not the whole interval. Dynamics of gaps provides boundaries for dynamics of points only up to the scale of the smallest gaps. Therefore, one could expect three regions in the level-spacing distribution picture: nonlinear non-smooth behavior in large $s$, power law in the middle-scales and breaking of power law in small-scales. This behavior is demonstrated in Fig. \ref{Fig:Logistic_LSD_twofigures}.
	
	It is wirthwhile to note that the reasoning outlined above is applicable for other polynomial-like dynamical systems since all of them can be approximated by piece-wise linear functions. If a piece-wise linear function has $m$ linear components with slopes $a_l$, then after finite number of iterations, in general case, there are $m$ intervals $\Delta_l$, which lengths is changed after an iteration only by multiplication of $a_l$. So, for large $n$ (i.e. small scales), $p_n(s)$ is a sum of multinomial distributions (for each interval $\Delta_l$):
	
	\begin{equation}
	p_n(s)=\sum_{l}\sum_{k_i} \frac{n!}{k_1!...k_m!} \delta(s-a^{k_1}_{1}\times\ldots \times a^{k_m}_{m}|\Delta_l|)
	\label{Eq:genLSD}
	\end{equation}
	
	If some of $a_l$ are close to each other then one obtains approximately the multinomial distribution with less number of variables. Let us assume that $a_l$ are distinguished enough. Then the main maximum for each interval $\Delta_l$ occurs at the point, where $a_l$ equal. In this case we obtain that $P(s)\sim\sum (s/\Delta_l)^{\ln m/\beta}$, where $\beta$ is the average of $a$'s.
	Therefore one can expect power-law behavior of level-spacing distribution for some kind of fractals.
	
	\section{Modified Sierpinski gasket}
	
	Next, we consider the level-spacing distributions of modified fractal Sierpinski gasket by using the exact diagonalization and some analytical estimations (an example of $3$ iterations of the fractal is shown in the Fig. \ref{fig:SGmod_cube}). The idea is to justify a hypothesis that the level-spacing distribution for fractals is asymptotically power function. Since the splitting of spectrum follows the power-law behavior, the idea is to force spectrum splitting by adding a parameter $\epsilon$, which is responsible for the hoppings between different congruent parts of the fractal. For example, we have some copies of $k-1$-th fractal iteration, then we glue them together to obtain the $k$-th iteration, but the hopping connecting these copies is $\epsilon^k$. Thus, with increasing $ k$ the copies of the fractal become asymptotically independent, and the energy spectrum splits by each iteration. For example, the hoping between sites $A$ and $B$ in Fig. \ref{fig:SGmod_cube} is $\epsilon^3$ and the hoping between sites $C$ and $D$ is $\epsilon^2$.
	
\begin{figure*}[ht!]
		\mbox{
		\includegraphics[width=0.4\linewidth]{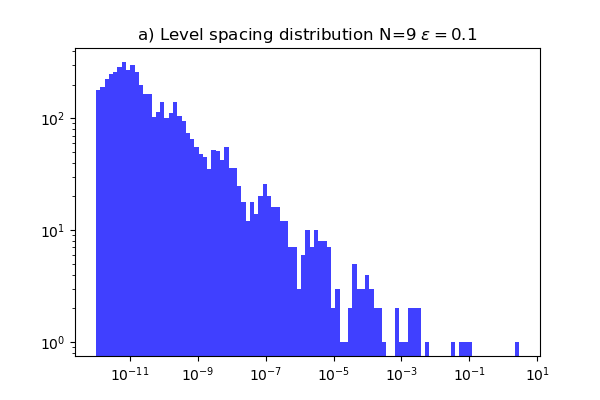}
		\includegraphics[width=0.4\linewidth]{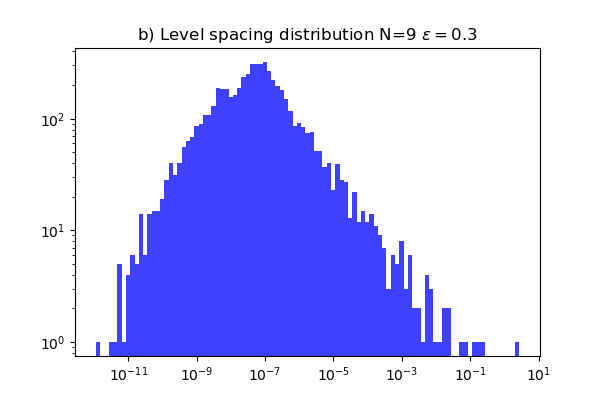}	
		}
	\mbox{
		\includegraphics[width=0.4\linewidth]{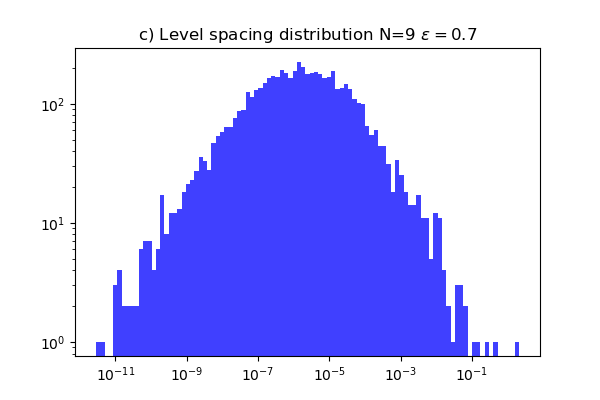}
		\includegraphics[width=0.4\linewidth]{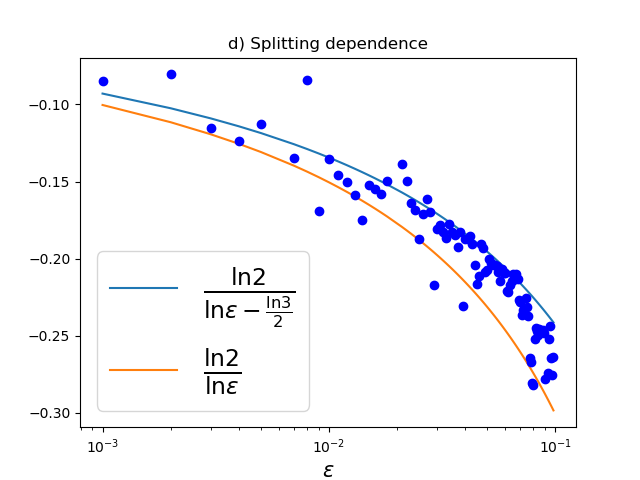}
	}
		\caption{\label{fig:SG_mod_pictures} (a), (b), (c) Level spacing distributions of modified Sierpinski gasket, $N$=9, $\epsilon$=0.1, 0.3 and 0.7 accordingly. (d) Dependence of the slope of power-law splitting of parameter $\epsilon$.}

	\end{figure*}

	For small iteration number $N$, linear dependence in log-log coordinates is not obvious, but with increasing $N$ such dependence becomes rather clear for modified Sierpinski gasket (Fig. \ref{fig:SG_mod_pictures}, these results are obtained numerically). For small $\epsilon$ linear dependence is also not obvious, however, the situation could be clearer with increasing $N$. The slope of level spacing distribution is increasing with increasing $\epsilon$.
	Near zero the slope changes sign, it can be connected with the finite number of iterations.
	
	One can check the idea of power-law splitting with the following estimation. Since the spectrum of Sierpinski gasket is obtained by a dynamical system with two branches, we can assume that the eigenvalues in modified gasket are also split into two at the next iteration. Therefore, the factor $m$ in the formula (\ref{Eq:genLSD}) should be equal to $2$. For small $\epsilon$, the characteristic size of splitting depends only on $\epsilon$. Then we obtain $P(s)\sim s^{\frac{\ln 2}{\ln \epsilon}}$. 

The numerical results are shown in Fig. \ref{fig:SG_mod_pictures}. The origin of corrected line in Fig. \ref{fig:SG_mod_pictures} is explained in the next Section. One can see that even this simple estimation describes the level-spacing distribution rather well.

	\section{Spectrum and paths}
	
	Let us consider the spectrum $\lambda_i$ of adjacency matrices $A$ of a finite realization of fractals. The spectrum is fully encoded in the statistical sum $Z(t)$:
	
	\begin{equation}
	Z(t)=\frac{1}{N}\sum_{i} e ^{\lambda_i t}
	\end{equation}
where $N$ is the matrix size and as a consequence, one has $Z(0)=1$.
	Self-similar matrices of different order are similar since their statistical sums are close to each other. In the case of self-similar matrix there is an hierarchy of embedded elements, which are isomorphic to each other on different scales.
	
	Statistical sum is connected to the traces of the adjacency matrix (the matrix shows which vertices are connected i.e. it is Hamiltonian matrix with hoppings equal to 1) powers by the formula:
	
	\begin{equation}
	Z(t)=\frac{1}{N} \Tr e^{t A}=\frac{1}{N}\sum\frac{t^n\Tr A^n}{n!}
	\end{equation}
	
	These traces are expressed by the total number of closed paths of length $n$ in a graph described by an adjacency matrix. Thus, the spectrum of a fractal is closely related to the geometry of paths on a fractal.
	
\begin{figure}[ht!]
		\mbox{
		\includegraphics[width=0.5\linewidth]{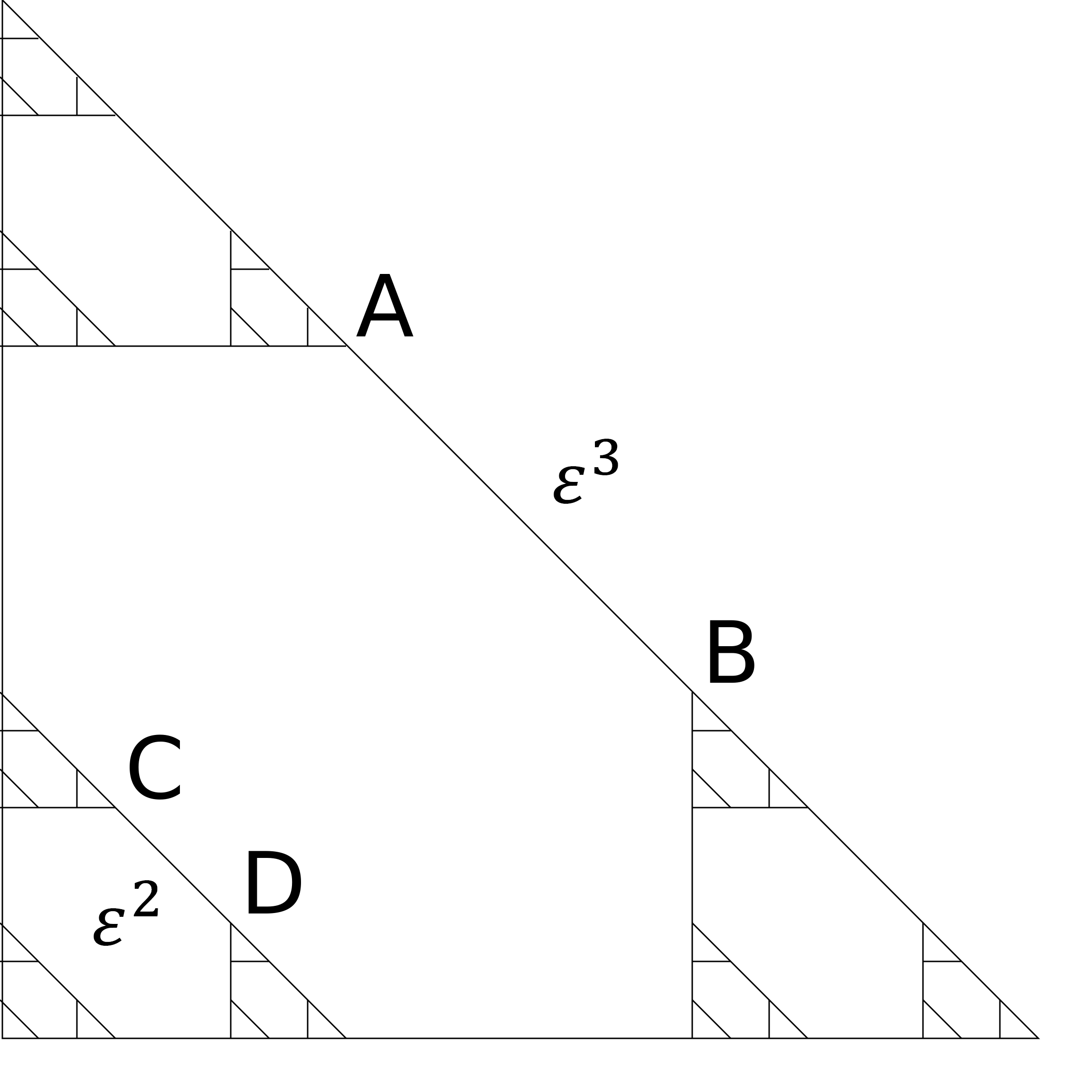}
		\includegraphics[width=0.5\linewidth]{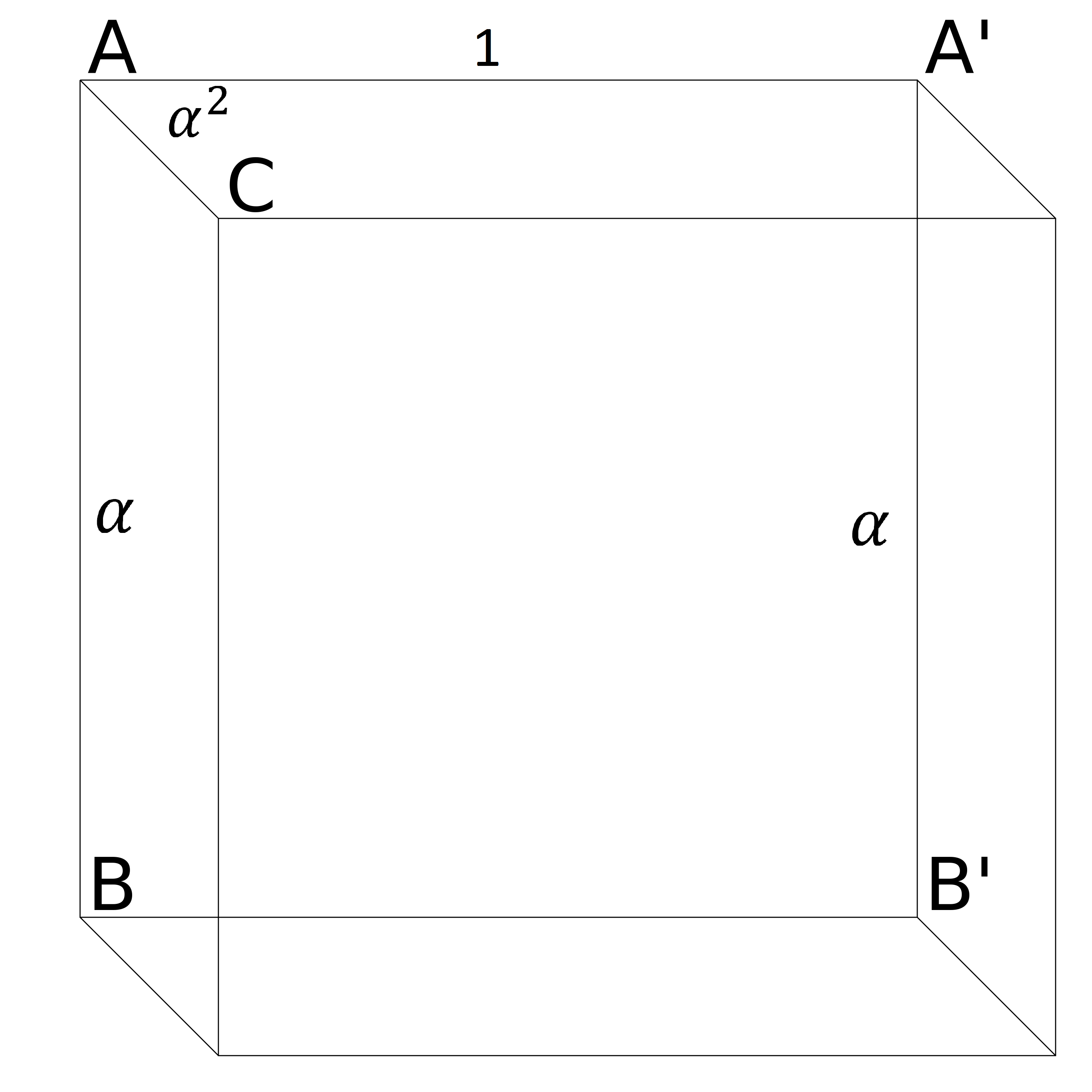}
		}
		\caption{\label{fig:SGmod_cube}Left: The modified Sierpinski gasket with 3 iterations. Right: An illustration to the example of a graph with Kantor set spectrum.}
	\end{figure}

	A simple example of a kind of fractal structure with power-law level-spacing statistics can be obtained by the following procedure. One starts with a basic figure, for example, a segment with a weight $\alpha$. At the first iteration, the vertices of a segment are themselves replaced by segments and corresponding points are connected with the weight $\alpha^2$. Then we have two copies of the square of previous iteration and connect corresponding point with the weight $\alpha^3$, i.e. on each iteration the vertices of the segment are replaced by a structure of previous iteration. So we obtain a n-dimensional cube with geometrically weighted edges. For example, in Fig. \ref{fig:SGmod_cube}, hoppings $AA'$ and $BB'$ equal to $1$, hoppings $AB$ and $A'B'$ equal to $\alpha$, hopping $AC$ is $\alpha^2$. If $n$ goes to infinity, the spectrum of such a system is a Kantor set consisting of points of the form $K={\pm\alpha\pm\alpha^2\pm\alpha^3\ldots}$. This spectrum corresponds to the limit set of the linear toy-model (Section \ref{Eq:toymodel}) with $\alpha=\beta$.
	
	The spectrum follows from the results described in Ref. [\onlinecite{Sayama2016}]. Let us consider the graph  $G\square H$ (Cartesian product of graphs) that obtained from two graphs $G$ and $H$ in the following way. One substitute graph $G$ into the vertices of graph $H$, i.e. there are copies of $G$ connected in the same way as the vertices in $H$. Each of connection consists of $N_G$ edges, where $N_G$ is number of vertices in graph $G$, and each of edges in a connection connects two corresponding vertices in two copies of graph $G$. Then the spectrum of a new graph $G\square H$ consists of the sums of eigenvalues of $G$ and $H$, $\sigma(G\square H)=\{\lambda_H+\lambda_G,\ \lambda_H \in \sigma(H),\ \lambda_G \in \sigma(G)\}$. This result can be shown by the following calculation:
	\begin{multline*}
	\Tr(G\square H)^n=\sum_{k} {n\choose k}\Tr G^k \Tr H^{n-k}=\\
	=\sum_{k}\sum_{i}\sum_{j}{n\choose k}\lambda^k_{Gi}\lambda^{n-k}_{Hj}=\sum_{i}\sum_{j}(\lambda_{Hj}+\lambda_{Gi})^n
	\end{multline*}
	In this iterative procedure the number of the connections from a copy of previous iteration to another copy growths exponentially with the increasing of iteration number. However, weighting in the geometric progression suppresses the growth of number of paths. For fractals with finite ramification number, the number of connections is finite on each iteration, therefore even without weighting one could expect similar behaviour of the spectrum.
	
	If there is no connection from one copy to the other, the spectrum is just degenerate, therefore only closed paths, which lie on a few copies, influence the spectrum. If on the $n$ iteration the number of connections from one copy of graph to another is $d^n$ (since the number of paths grows exponentially), then the number of influencing closed paths is proportional to $d^n$. Since with weighting $\alpha^n$ there is a splitting of spectrum proportional to $\alpha^n$, then with finite connection and without weighting  one could expect the average splitting to be proportional to $d^{-n}$ on the $n$ iteration. This gives a power-law level-spacing distribution of the spectrum.
	
	Let us describe a rough estimation for the simplest case, when there are two copies of a graph with adjacency matrix $A$, which are connected by an edge with hopping $\alpha$ at the corresponding points. Having assumed that eigenvalues $\lambda_j$ are splitting by the same average value $\delta\lambda$, we obtain:
	
	\begin{equation*}
	2\Tr A^2 + 2\alpha^2 = \sum_{j}(\lambda_j+\delta\lambda)^2+\sum_{j}(\lambda_j-\delta\lambda)^2
	\end{equation*}
	
	The left side is the trace of the adjacency matrix of the whole system. Thus, $\delta\lambda\sim\alpha/\sqrt{N}$, where $N$ is the number of eigenvalues. Since the size of a fractal increases geometrically with iterations, splitting between eigenvalues is also exponential. This estimation also leads to the correction in the dependence of the splitting slope in Fig. \ref{fig:SG_mod_pictures}. Since $N$ on each iteration of modified gasket increases $3$ times and weights multiply by $\epsilon$, effective splitting should be proportional to $\epsilon/\sqrt3$. One can see that corrected line fits calculated slopes better.
	
	It also interesting to note that the difference between fractals with different ramification number becomes clearer by this approach since their statistical sums have different convergent properties.
	
	\section{Summary}
	
	In this work the power-law spectrum statistics is demonstrated for some fractal structures and the explanations of this phenomena are proposed.
	
	 The first approach is from the point of view of the limit set of dynamical systems. It was shown that if the spectrum of a fractal can be obtained as a limit set of a smooth dynamical system, then the level-spacing distribution of the spectrum is asymptotically the sum of power-law distributions. The calculations were numerically checked for the simple model of piece-wise linear function.
	
	The second approach is connected with the geometry of the paths in fractals.
	The idea is that the hierarchical structure of a fractal induces the hierarchical structure of the number of closed paths, which in turn induces a splitting of the spectrum of a fractal in each iteration. This part is more vague, but, nevertheless, one can estimate the slope of the power-law distributions, which fits the results obtained from the numerical calculations.
	
	We conclude that the power-law of the level-spacing-distribution can be a general feature of fractals, which is differently from that of disordered systems and they constitute a separate class of systems.
	
	\section*{Acknowledgements}
	
	This work was supported by the National Science Foundation
	of China under Grant No. 11774269 and by the Dutch
	Science Foundation NWO/FOM under Grant No. 16PR1024
	(S.Y.), and by the European Research Council Advanced
	Grant program (contract 338957) (M.I.K.). Support by the
	Netherlands National Computing Facilities foundation (NCF),
	with funding from the Netherlands Organisation for Scientific
	Research (NWO), is gratefully acknowledged.
		
\bibliographystyle{apsrev}

\begin{thebibliography}{}
\bibitem{Dyson1962}
F.~J. {Dyson}, ``Statistical theory of the energy levels of complex systems.
i,'' {\em Journal of Mathematical Physics}, vol.~3, pp.~140--156, 1962.

\bibitem{Peerenboom1981}
J.~A.~A.~J. {Perenboom}, P.~{Wyder}, and F.~{Meier}, ``Electronic properties of
small metallic particles,'' {\em Physics Reports}, vol.~78, pp.~173--292,
1981.

\bibitem{Halperin1986}
W.~P. Halperin, ``Quantum size effects in metal particles,'' {\em Rev. Mod.
	Phys.}, vol.~58, pp.~533--606, Jul 1986.

\bibitem{Stockmann_2000}
H.~J. Stockmann, {\em Quantum Chaos: An Introduction}.
\newblock Cambridge University Press, Cambridge, 2000.

\bibitem{Bohr_1969}
A.~Bohr and B.~Mottelson, {\em Nuclear Structure, Vol 1: Single-Particle
	Motion}.
\newblock W. A. Benjamin, New York, 1969.

\bibitem{Beenakker1997}
C.~W.~J. Beenakker, ``Random-matrix theory of quantum transport,'' {\em Rev.
	Mod. Phys.}, vol.~69, pp.~731--808, Jul 1997.

\bibitem{Mehta_2004}
M.~L. Mehta, {\em Random Matrices}.
\newblock Elsevier, Amsterdam, 2004.

\bibitem{HavAbr1987}
S.~Havlin and D.~Ben-Avraham, ``Diffusion in disordered media,'' {\em Advances
	in Physics}, vol.~36, no.~6, pp.~695--798, 1987.

\bibitem{EversMirlin2008}
F.~Evers and A.~D. Mirlin, ``Anderson transitions,'' {\em Rev. Mod. Phys.},
vol.~80, pp.~1355--1417, Oct 2008.

\bibitem{Polini_etal2013}
M.~Polini, F.~Guinea, M.~Lewenstein, H.~C. Manoharan, and V.~Pellegrini,
``Artificial honeycomb lattices for electrons, atoms and photons,'' {\em
	Nature Nanotechnology}, vol.~625, 2013.

\bibitem{Gibertini_etal2009}
M.~Gibertini, A.~Singha, V.~Pellegrini, M.~Polini, G.~Vignale, A.~Pinczuk,
L.~N. Pfeiffer, and K.~W. West, ``Engineering artificial graphene in a
two-dimensional electron gas,'' {\em Phys. Rev. B}, vol.~79, p.~241406, Jun
2009.

\bibitem{Shang_etal2015}
J.~Shang, Y.~Wang, M.~Chen, J.~Dai, X.~Zhou, J.~Kuttner, G.~Hilt, X.~Shao,
J.~M. Gottfried, and K.~Wu, ``Assembling molecular sierpiński triangle
fractals,'' {\em Nature Chemistry}, vol.~7, 2015.

\bibitem{Hohlfeld2011}
R.~Hohlfeld and N.~Cohen, ``Self-similarity and the geometric requirements for
frequency independence in antennae,'' {\em Fractals}, vol.~07, 11 2011.

\bibitem{Huang2010}
X.~Huang, S.~Xiao, D.~Ye, J.~Huangfu, Z.~Wang, L.~Ran, and L.~Zhou, ``Fractal
plasmonic metamaterials for subwavelength imaging,'' {\em Opt. Express},
vol.~18, pp.~10377--10387, May 2010.

\bibitem{Tosatti_1986}
L.~Pietronero and E.~T. (Editors), {\em Fractals in Physics}.
\newblock Elsevier, Amsterdam, 1986.

\bibitem{Feder_1988}
J.~Feder, {\em Fractals}.
\newblock Plenum Press, New York, 1988.

\bibitem{Strichartz2006}
R.~Strichartz, {\em Differential analysis on fractals: a tutorial}.
\newblock Princeton University Press, 2006.

\bibitem{Brzezetal2018}
M.~Brzezinska, A.~M. Cook, and T.~Neupert, ``Topology in the
sierpinski-hofstadter problem,'' 2018.

\bibitem{KosKrzys2017}
A.~Kosior and K.~Sacha, ``Localization in random fractal lattices,'' {\em Phys.
	Rev. B}, vol.~95, p.~104206, Mar 2017.

\bibitem{Veen2016}
E.~van Veen, S.~Yuan, M.~I. Katsnelson, M.~Polini, and A.~Tomadin, ``Quantum
transport in sierpinski carpets,'' {\em Phys. Rev. B}, vol.~93, p.~115428,
Mar 2016.

\bibitem{Veen2017}
E.~van Veen, A.~Tomadin, M.~Polini, M.~I. Katsnelson, and S.~Yuan, ``Optical
conductivity of a quantum electron gas in a sierpinski carpet,'' {\em Phys.
	Rev. B}, vol.~96, p.~235438, Dec 2017.

\bibitem{Westerhout2018}
T.~Westerhout, E.~van Veen, M.~I. Katsnelson, and S.~Yuan, ``Plasmon
confinement in fractal quantum systems,'' {\em Phys. Rev. B}, vol.~97,
p.~205434, May 2018.

\bibitem{Hernando2015}
A.~Hernando, M.~Sulc, and J.~Vanicek, ``Spectral properties of electrons in
fractal nanowires,'' 2015.

\bibitem{KatEvan1996}
G.~N. Katomeris and S.~N. Evangelou, ``Level statistics for electronic states
in a disordered fractal,'' {\em Journal of Physics A: Mathematical and
	General}, vol.~29, no.~10, p.~2379, 1996.

\bibitem{Teplyaev_2007}
A.~Teplyaev, ``Spectral zeta functions of fractals and the complex dynamics of
polinomials,'' {\em Transactions of the American Mathematical Society},
vol.~359, p.~4339–4358, 2007.

\bibitem{Sabot_2003}
C.~Sabot, ``Spectral properties of self-similar lattices and iteration of
rational maps,'' {\em Memoires de la SMF}, vol.~92, 2003.

\bibitem{Domany_1983}
E.~{Domany}, S.~{Alexander}, D.~{Bensimon}, and L.~{Kadanoff}, ``Solutions to
the schrodinger equation on some fractal lattices,'' {\em Phys. Rev. B},
vol.~28, pp.~3110--3123, 1983.

\bibitem{Strogatz_2000}
S.~Strogatz, {\em Nonlinear Dynamics And Chaos}.
\newblock CRC Press, 2000.

\bibitem{Sayama2016}
H.~Sayama, ``Estimation of laplacian spectra of direct and strong product
graphs,'' {\em Discrete Applied Mathematics}, vol.~205, pp.~160 -- 170, 2016.
\end{thebibliography}

\end{document}